\documentclass{article}
\usepackage{epsfig}
\begin{document}

\title{The Importance of $\pi N \rightarrow K\Lambda$ Process for the Pole Structure of the P11 Partial Wave T-matrix
in the Coupled-Channel Pion-Nucleon Partial Wave Analysis\footnote{\uppercase{T}he complete set of transparencies can be found at \newline http://hadron.physics.fsu.edu/\uppercase{NSTAR}2005/\uppercase{TALKS/F}riday/\uppercase{P}arallel\_\uppercase{C/Z}auner.pdf}}

\author{B. Zauner, S. Ceci and A.~\v{S}varc \\ \\ \emph{Rudjer Boskovic Institute} \\ \emph{Bijeni\v{c}ka Cesta 54}\\ \emph{10000 Zagreb, Croatia} \\ \emph{E-mail: Branimir.Zauner@irb.hr}}

\maketitle

\section{Introduction}
The pole structure of P11 partial wave in pion-nucleon scattering still has issues. Different analyses give different number of poles and different pole positions. The analysis that is most widely used (VPI/GWU)\cite{Arn04}, reports a single resonant pole in the first Riemann sheet. We will show that this analysis does not exclude other poles, namely poles in W=1700 MeV region, and that they appear the moment we include the $\pi N \rightarrow K\Lambda $ inelastic channel.

\section{Tools and data sets}

\textbf{Tools}

We are using a coupled-channel CMU-LBL type formalism. For the collection of formulae we refer the reader either to
original paper by Cutkosky et.al\cite{Cut79}  or to one of the more recent CC\_PWAs; Zagreb\cite{Bat98} or Pittsburgh/ANL\cite{Vra00}.
\\
\noindent
\textbf{Data Sets}

For elastic channel, we used single energy solutions from SAID site\cite{SAID}. For the $\pi N \rightarrow K\Lambda$ channel, we used T matrix element we obtained in a single channel partial wave analysis from experimental data.
We fitted data to three partial waves - S11, P11 and P13. We had a single resonance per wave, except in P11, where
we allowed for two. Error bars were put a posteriori, in order to make statistical weight of this partial wave data smaller.

\begin{figure}[!ht]
\centerline{\epsfxsize=3in\epsfbox{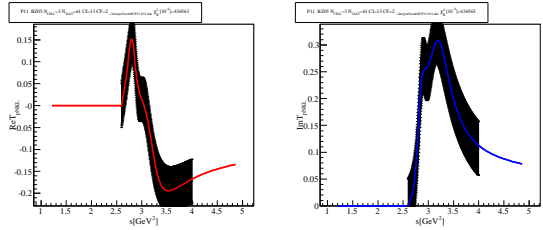}}
\caption{The $\pi N \rightarrow K\Lambda$ P11 partial wave T matrix element.}
\end{figure}

\section{Poles around 1700 MeV are a direct consequence of inelastic channels}

\subsection{Fitting only FA02}
First, we shall show that poles around 1700 MeV are not artefacts of the CMU-LBL model. We fitted only FA02 SES (and had a free unitarizing channel to preserve unitarity) and found out that a single pole is sufficient to describe the data well. There is no sign of any pole structure in the 1700 MeV region.

\begin{figure}[!ht]
\centerline{\epsfxsize=2.25in\epsfbox{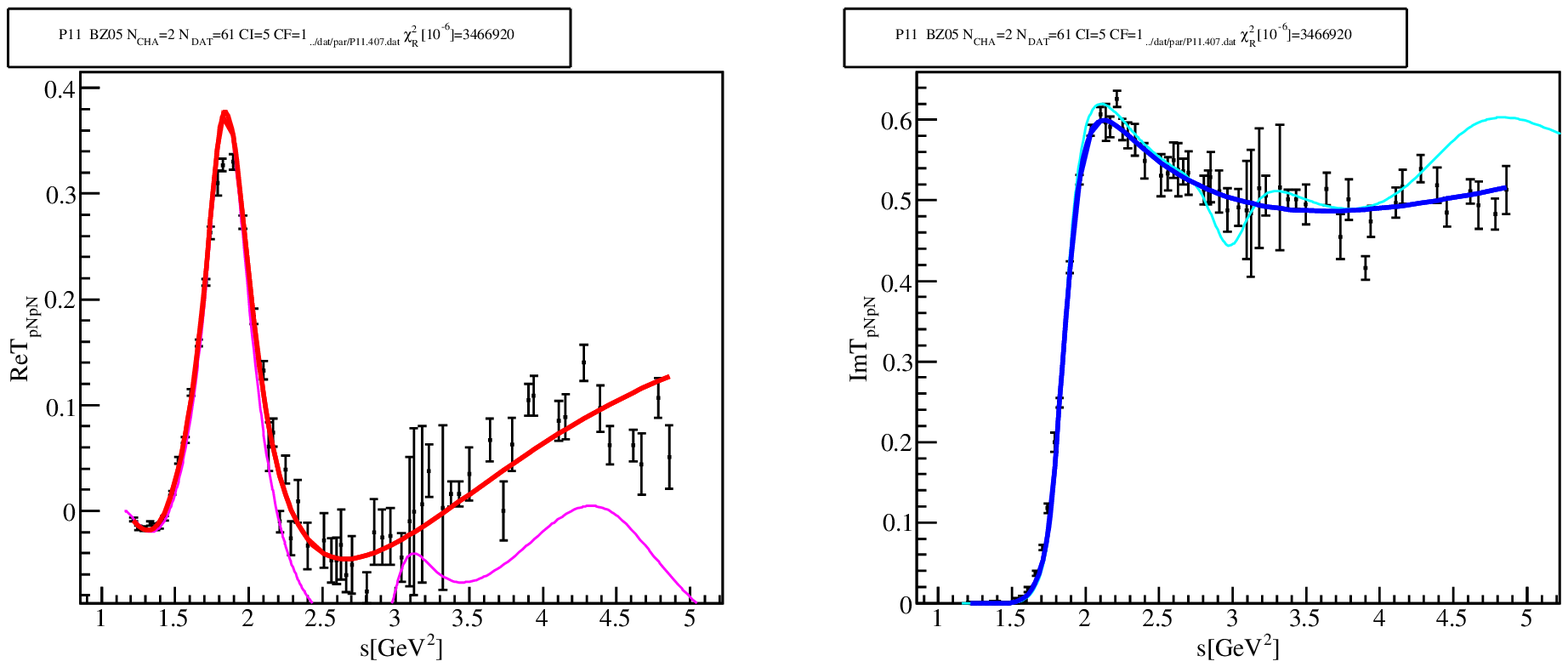} \epsfxsize=2.25in\epsfbox{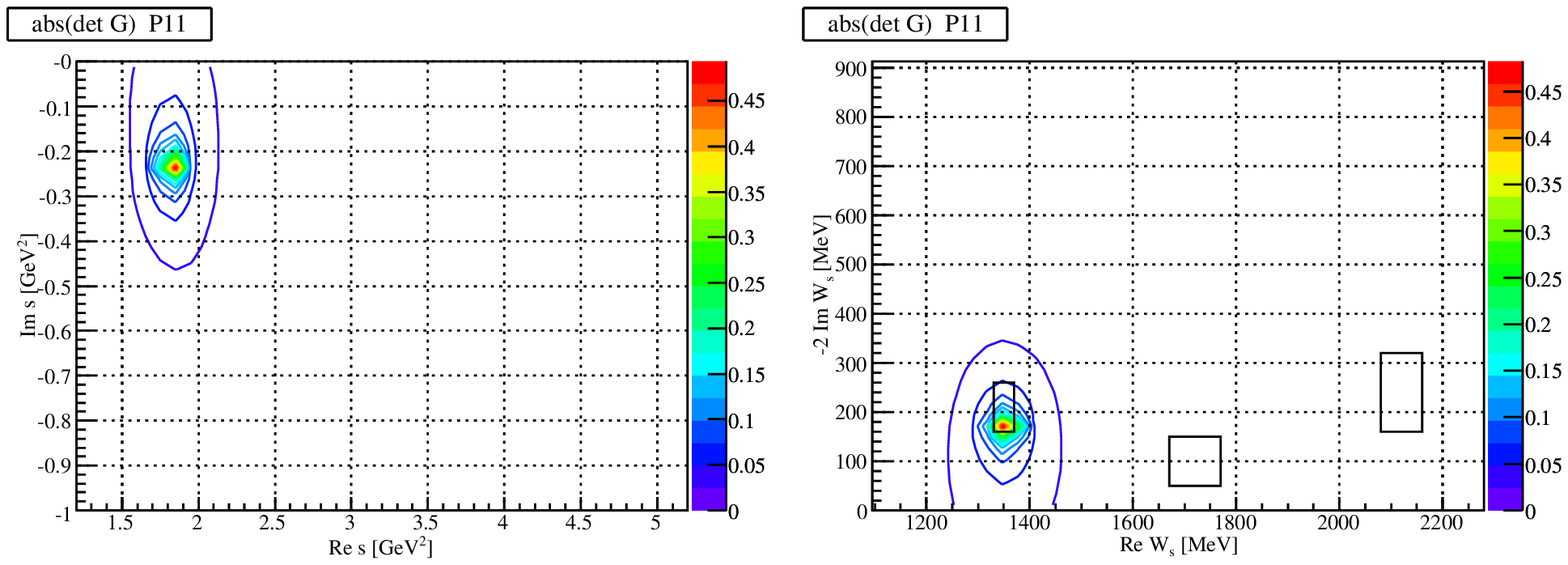}}
\caption{Result for FA02 with a single pole in the resonance region. Two right graphs show pole positions, where one can see good agreement with Particle Data Group pole positions (black rectangles on the picture)}
\end{figure}

If we add another pole, agreement with the data is slightly improved in the high energy region, as seen in fig.3.
\begin{figure}[!ht]
\centerline{\epsfxsize=2.25in\epsfbox{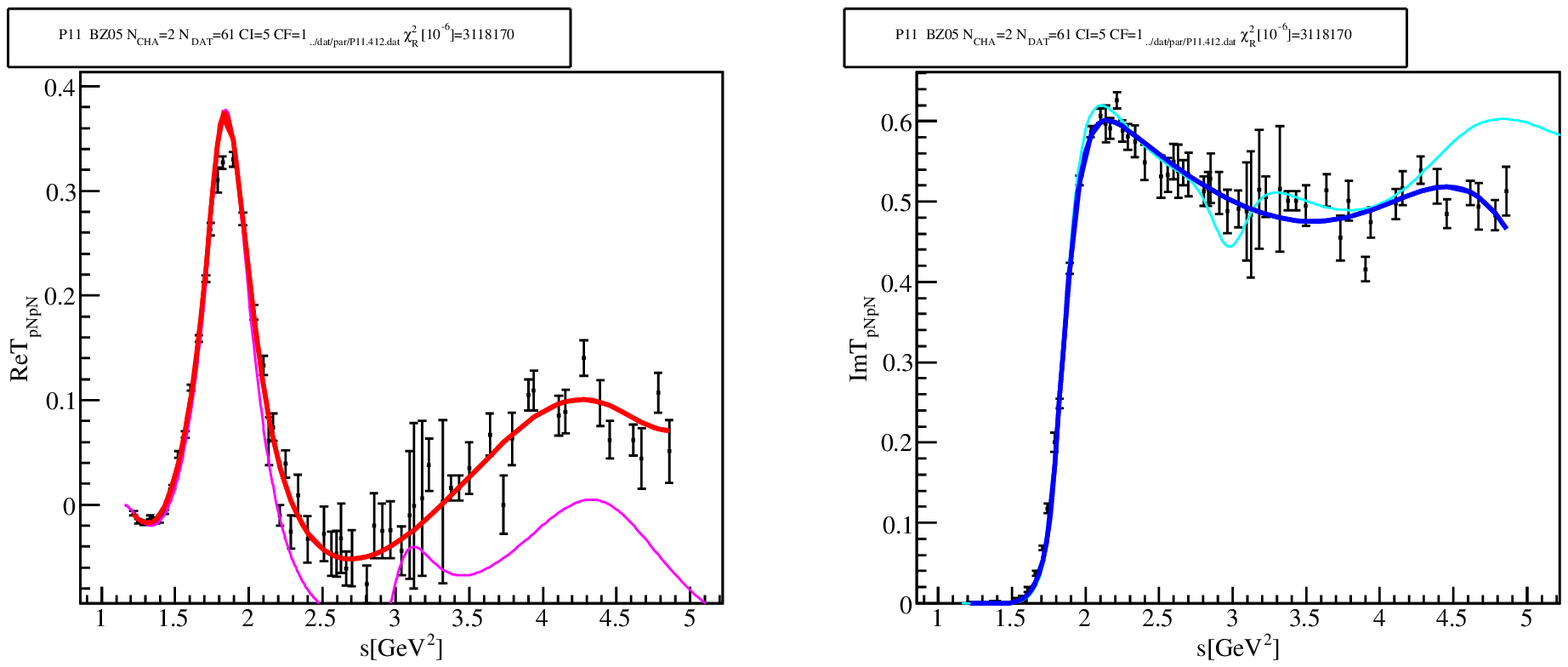} \epsfxsize=2.25in\epsfbox{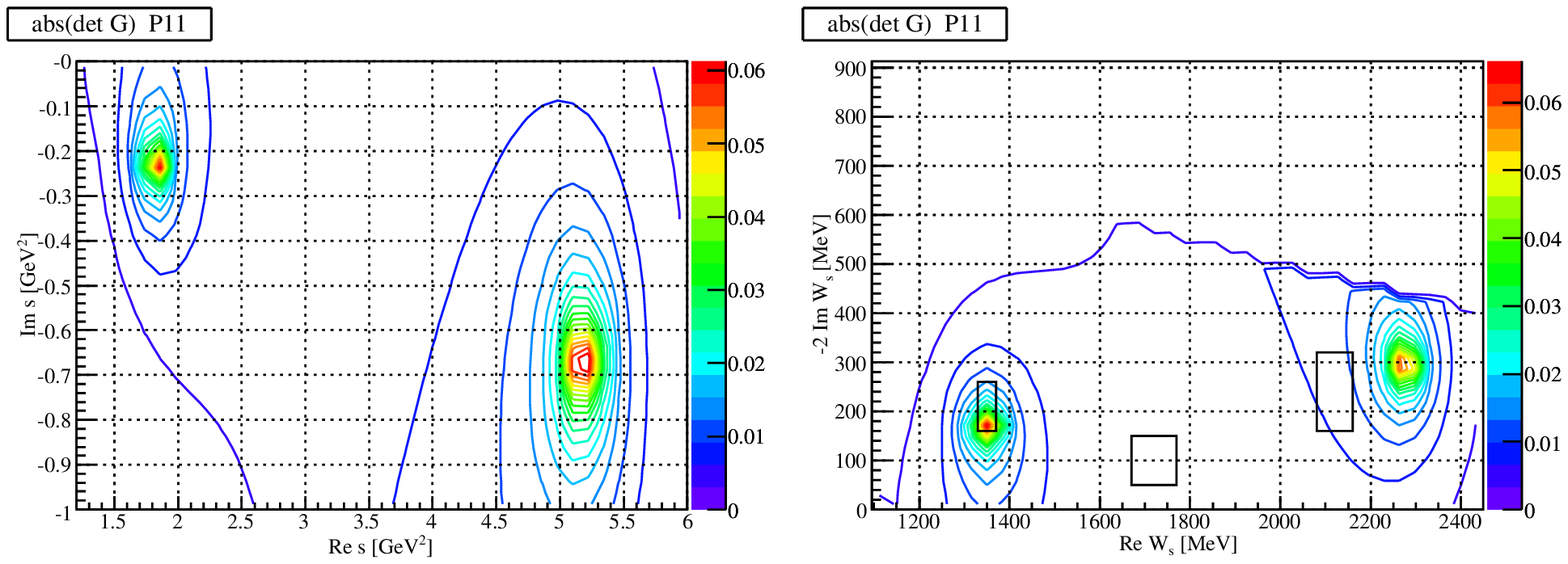}}
\caption{Result for FA02 with two poles in the resonance region.}
\end{figure}
In the right part of fig.3, one can see that Roper resonance did not move, and that we got another resonance in 2300 MeV region, far above the region we are interested in. Therefore we conclude that VPI/GWU SES do not need a pole in the 1700 MeV region. Consequently, there is no need for the N(1710)P11 resonance.

\pagebreak
\subsection{Including the $\pi N \rightarrow K\Lambda$ channel }
When we include the $\pi N \rightarrow K\Lambda$ T matrix (now we have three channels to fit - elastic, inelastic and unitarizing), we see that a single pole in the resonance region miserably fails to reproduce the data. Moreover, in CMU-LBL model, we have two poles in the subthreshold region that are simulating background. Inclusion of the inelastic channel (when we put a single pole in the resonance region) forces them to simulate real resonances, as one can see in fig.4.

\begin{figure}[!ht]
\centerline{\epsfxsize=2.25in\epsfbox{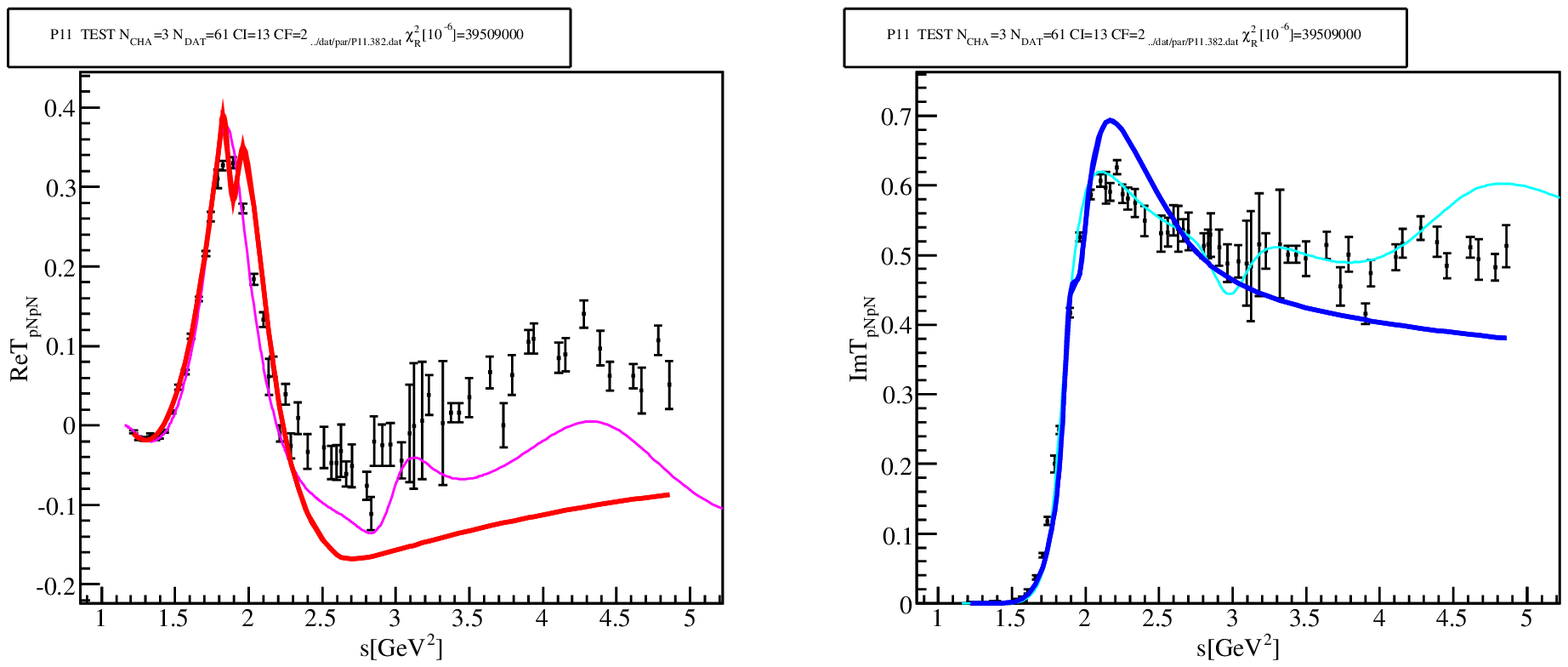} \epsfxsize=2.25in\epsfbox{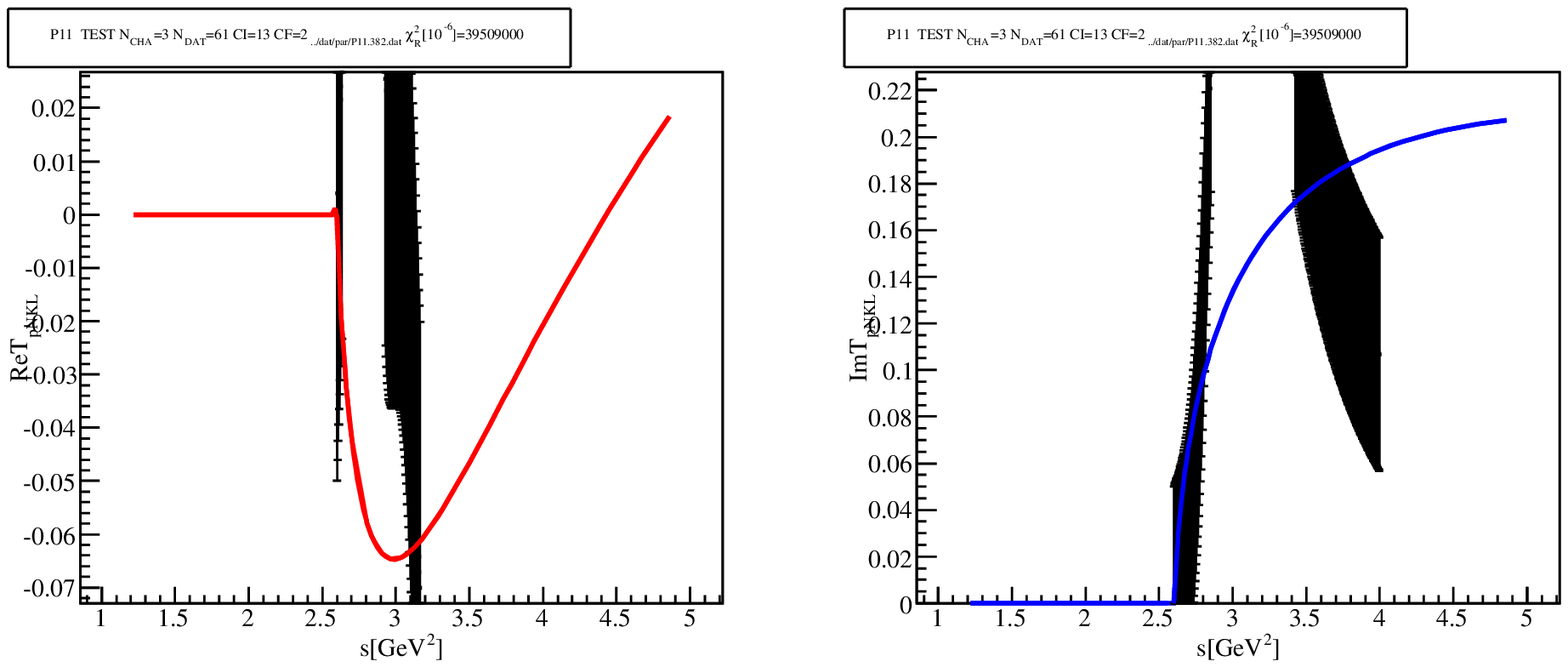}}
\centerline{ \epsfxsize=3in\epsfbox{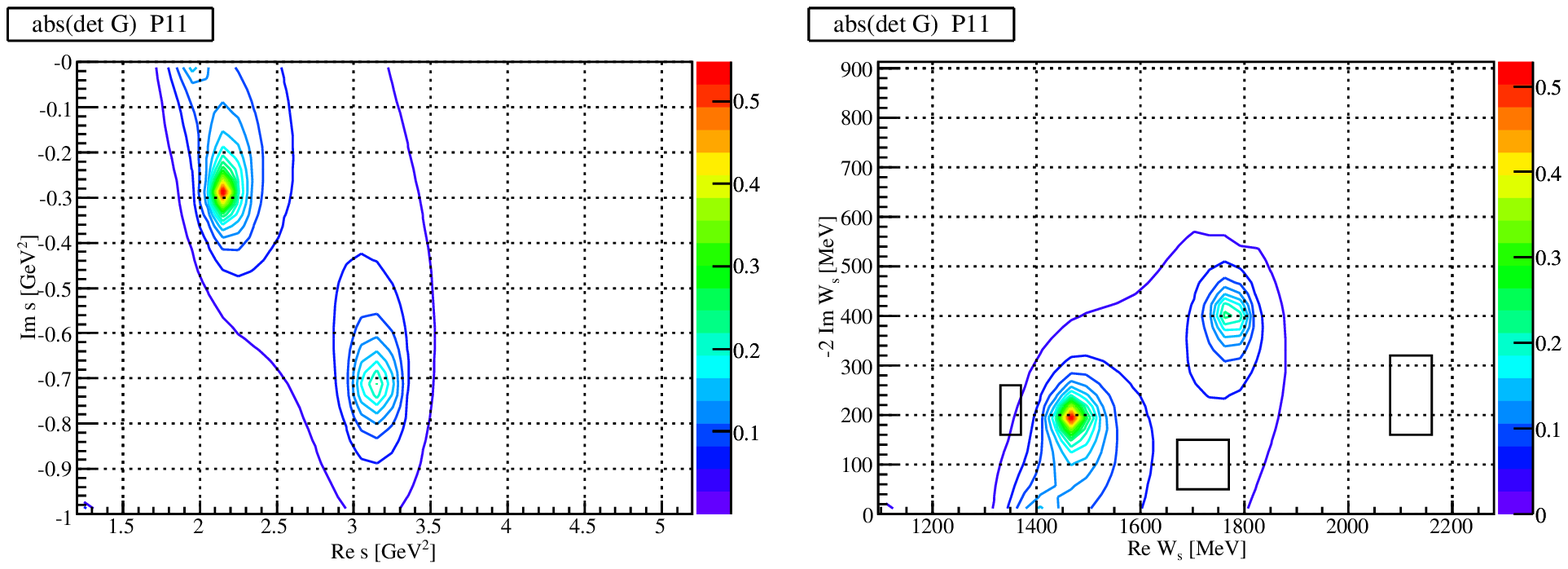}}
\caption{Result for FA02 SES and $\pi N \rightarrow K\Lambda$ T matrix element. We used a single pole in resonant region.}
\end{figure}

After further fitting, and trying various number of resonances, first number of poles in the resonant region that reproduces both data sets well and does not force the background poles into resonant region is four. In the fig.4 we can see that we have pole structure at around 1700 MeV, i addition to the Roper resonance pole.

\begin{figure}[!ht]
\centerline{\epsfxsize=2.25in\epsfbox{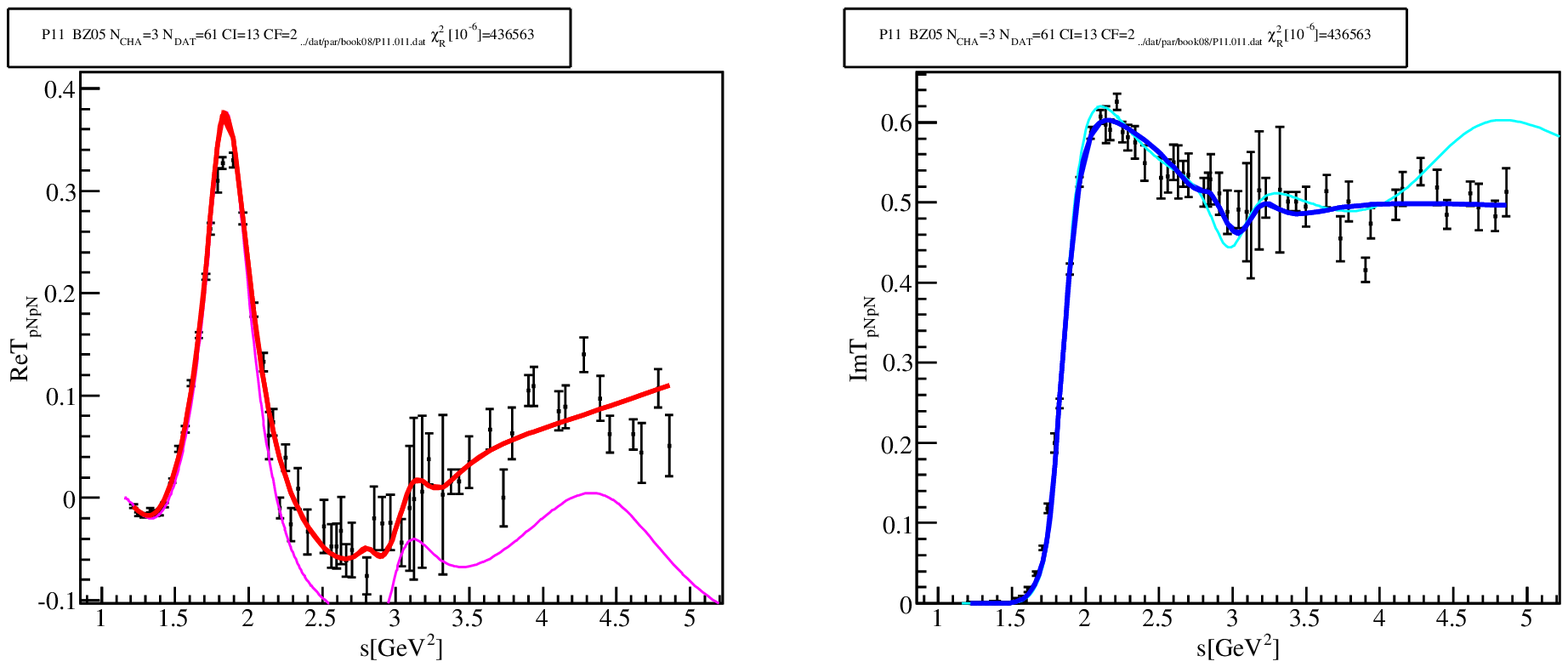} \epsfxsize=2.25in\epsfbox{5kl.eps}}
\centerline{\epsfxsize=4in\epsfbox{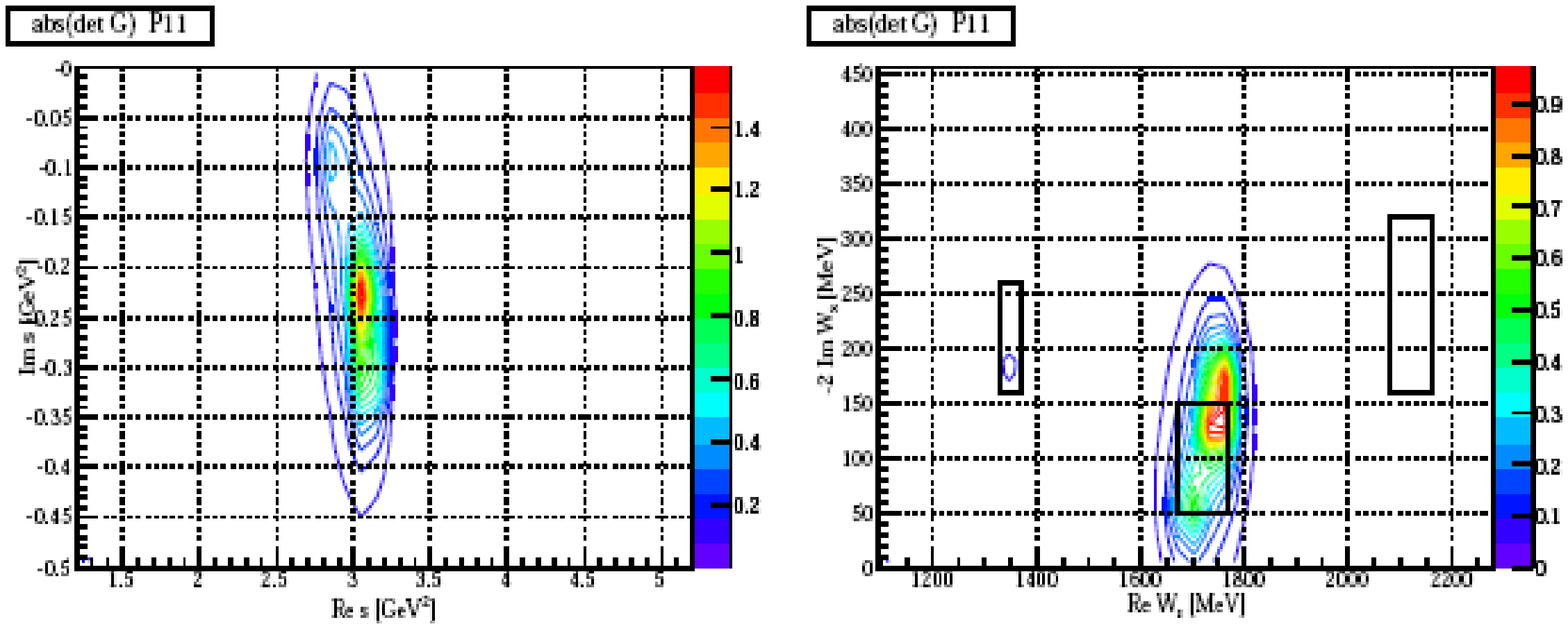}}
\caption{Result for FA02 SES and $\pi N \rightarrow K\Lambda$ T matrix element. We used 4 poles in resonant position.}
\end{figure}

\section{Conclusions}
Procedure that was presented here points out that the lack of evidence for N(1710)P11 in many recent analyses is due to the single channel treatment. Introduction of inelastic channels (namely, $\pi N \rightarrow K\Lambda$) is crucial to the confirmation of the existence of N(1710)P11.


\begin{thebibliography}{0}

\bibitem{Arn04}R. A. Arndt et. al.,  Phys. Rev. {\bf C69}, 035213 (2004).

\bibitem{Cut79} R.E. Cutkosky, C.P. Forsyth, R.E. Hendrick and R.L. Kelly,  Phys. Rev. {\bf D20}, 2839 (1979).

\bibitem{Bat98}  M. Batini\'{c}, I. \v{S}laus, A. \v{S}varc and B.M.K. Nefkens,  Phys. Rev. {\bf C51}, 2310 (1995);
  M. Batini\'{c}, I. \v{S}laus, A. \v{S}varc, B.M.K. Nefkens and T.S.-H. Lee,  Physica Scripta {\bf 58}, 15 (1998).

\bibitem{Vra00}  T.P. Vrana, S.A. Dytman and T.S.-H- Lee,  Phys. Rep. {\bf 328}, 181 (2000).

\bibitem{SAID} see: http://gwdac.phys.gwu.edu/

\bibitem{PDG}  S. Eidelman et al., Physics Letters B592, 1 (2004) 

\end{thebibliography}
\end{document}